\documentclass[prx,letterpaper,aps,10pt,superscriptaddress,twocolumn,floatfix,showpacs]{revtex4-1}
\usepackage{graphicx}
\usepackage{amsmath}
\usepackage{amsfonts}
\usepackage{float}
\usepackage{amssymb}
\usepackage{epsfig}
\usepackage{epstopdf}
\DeclareGraphicsExtensions{.pdf,.eps,.png,.jpg,.mps}
\usepackage[pdftex]{color}
\usepackage{amsmath,graphicx,amssymb,braket,xcolor,subfigure,upgreek}
\usepackage[colorlinks, linkcolor=blue, citecolor=blue, urlcolor=blue, breaklinks=true]{hyperref}
\usepackage{microtype}
\usepackage{bbm}
\usepackage{color}
\usepackage{lineno}

\newcommand{\non}{\nonumber}

\begin{document}
\title{Continuous-Variable Quantum Teleportation Using Microwave Enabled Plasmonic Graphene Waveguide}
\author{Muhammad Asjad}
\author{Montasir Qasymeh}
\affiliation{ Electrical and Computer Engineering Department, Abu Dhabi University, 59911 Abu Dhabi, UAE}
\author{Hichem Eleuch}
\affiliation{ Department of Applied Physics and Astronomy, University of Sharjah, Sharjah, UAE}

\affiliation{ Institute for Quantum Science and Engineering, Texas AM University, College Station, TX 77843 USA}


\begin{abstract}
We present a scheme to generate continuous variable bipartite entanglement between two optical modes in a hybrid optical-microwave-plasmonic graphene waveguide system. In this scheme, we exploit the interaction of two light fields coupled to the same microwave mode via Plasmonic Graphene Waveguide to generate two-mode squeezing, which can be used for continuous-variable quantum teleportation of the light signals over large distances. Furthermore, we study the teleportation fidelity of an unknown coherent state. The teleportation protocol is robust against the thermal noise associated with the microwave degree of freedom.  
\end{abstract}


\maketitle
\section{ Introduction}

The transfer of quantum states between distant nodes of a quantum network is a basic task for quantum information processing. All protocols used for quantum states transmission require nonlocal correlations, or entanglement, between the sender and the receiver systems.
Entanglement is at the basis of quantum mechanics and almost every quantum information protocol. Recent progress in the field of quantum communication and computation has shown that entanglement is a prominent resource for many applications in quantum technology \cite{MA}.
Entanglement can be exploited to perform various quantum information tasks such as quantum key distribution\cite{ Ekert,Waks}, quantum cryptographic protocols \cite{Benn}, dense coding \cite{Bennett, Braunstein} and quantum teleportation \cite{Ben,Lev,Brau}.
Entanglement has been investigated in several experimental works in different plateforms such as entanglement between two optical fields using a beam splitter \cite{Kim, Ber} or a nonlinear medium \cite{Lukin,ABM}, entanglement between two trapped ions \cite{Turch}, generation of an entangled photon-phonon pairs \cite{Yang}, entanglement between distant optomechanical systems \cite{asjadsq,Korppi, Joshi,Raymond}, and entanglement in waveguide quantum electrodynamics systems \cite{zhang1,zhang2,zhang3}, to mention few examples. 

Generating entanglement between microwave (optical) and optical fields is very crucial for the combination of superconductivity with quantum photonic systems \cite{Fan}, which is very important for efficient quantum computation and communication. Optomechanical systems can generate entangled pairs of optical and (optical) microwave fields \cite{Sbam, Sban}. However, the entanglement of the microwave and optical photons using mechanical resonators is very sensitive to the thermal noise. In earlier works \cite{Qas,Qass, Qp}, two authors of our group have proposed a scheme for an efficient low noise conversion of quantum electrical signal to optical signal and vice versa by using many layers of graphene. It has been shown that a few microvolts driving quantum microwave signal can be efficiently converted to the optical frequency domain by exploiting several graphene layers. 

Two-mode squeezed light is the primary resource for quantum information and quantum communication with continuous variable systems \cite{Bra, Ferraro, Adesso, Weedbrook}. A hybrid two-mode squeezing of microwave and optical fields has also been recently proposed by our group using a plasmonic graphene-loaded capacitor \cite{Qass}. In this paper, we develop such a scheme to realize  continuous variable (i.e., CV) entangled states between two different radiations at different wavelengths by using a superconducting electrical capacitor which is loaded with graphene plasmonic waveguide and driven by a microwave quantum signal. The interaction of the microwave mode with the two optical modes is used for the generation of stationary entanglement between the two output optical fields. We then illustrate a practical use of the resulting CV-entangled state to teleport an unknown coherent state over a long distance with high efficiency. The stationary entanglement and the quantum teleportation fidelity are found robust with respect to the thermal microwave photons that are associated with the microwave degree of freedom.

This paper is organized as follows. In Section \ref{sec2},  we introduce the system and derive the governing quantum Langevin equations. In Section \ref{sec3}, the generation of stationary output entanglement between the two optical modes is investigated in details. 
In Section \ref{sec4}, we show that the stationary output entanglement between the two optical modes can be used for teleportation of an unknown coherent state with high efficiency. In the last Section \ref{sec5}, we draw the conclusion remarks.
\section{system} \label{sec2}
\begin{figure}[tb]
\centering
\includegraphics[width=1\columnwidth]{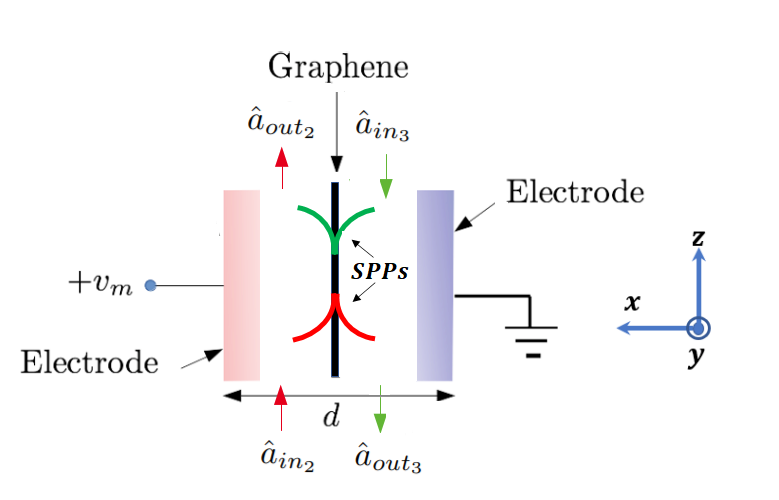}
\caption{The proposed superconducting electrical capacitor driven by microwave signal $v_m$ and  loaded with plasmonic graphene waveguide.}
\label{figsch}
\end{figure}

As shown in Fig. \ref{figsch}, we consider a quantum microwave signal  $v_m =v e^{-i \omega_m t}+c.c$ with frequency $\omega_m$  driving a superconducting capacitor of capacitance $\mathcal{C}=\epsilon \epsilon_0/d$. Here, $d$ is the distance between the two plates of the capacitor. 
We consider the two plates lying in the \textit{yz-plane} (with width $W$ along the $y$-dimension and length $L$ along the $z$-dimension) while a single graphene layer is placed at $z=0$. In addition to the microwave biasing, optical fields are launched to the graphene waveguide creating surface plasmon polariton (i.e., SPP) modes.
The SPP modes can be coupled to (and out of) the graphene waveguide using a nano-coupler or a tapered waveguide. Another possible way is to utilize a
nano-laser sources that are integrated and coupled to the graphene waveguide \cite{Qass2}.

The electrical and magnetic fields associated with a surface plasmon polariton mode, of a frequency   $\omega$, propagating along such a graphene waveguide are given by  $\vec{\mathcal{E}}=\mathcal{U}(z)\left(\mathcal{D}_x(x) \vec{e}_x  +\mathcal{D}_z(x) \vec{e}_z\right) e^{-i(\omega t-\beta z)}+c.c.$, 
and $\vec{\mathcal{H}}=\mathcal{U}(z)\mathcal{D}_y(x) \vec{e}_y  e^{-i(\omega t-\beta z)}+c.c.$, respectively. Here, $\mathcal{U}(z)$ is the complex amplitude, $\mathcal{D}_{i}(x) =  i \mathcal{K}_{i} /\omega \epsilon\epsilon_0\left\{ e^{\alpha x}\, \text{for} \, x<0;  e^{-\alpha x} \, \text{for} \, x>0,\right\}$ is the spatial distribution of the surface plasmon polariton mode for $i=x,y,z$, $\mathcal{K}_{x}=\beta$, $\mathcal{K}_{y}= -i \omega \epsilon\epsilon_0$, and $\mathcal{K}_{z}=\alpha$. Here, $\alpha=\sqrt{\beta^2-\epsilon \omega /c}$ is the transverse decaying factor in the $x$-direction around the graphene layer and $\beta=\omega /c\sqrt{1-\dfrac{2}{Z_0\zeta}}$ is the propagation constant with $c$ is the speed of light, and $Z_0$ represents the free space impedance. Importantly, in this work the spacing between the two plates is considered much greater than the reciprocal of the transverse decay factor (i.e., $d \gg \frac{1}{\alpha}$). It then follows that the plates are not impacting the propagation of the SPP modes. Nevertheless, the microwave field is interacting with the SPP modes through electrically modulating the graphene conductivity as will be shown in the following.

The graphene waveguide can be characterized by its conductivity $\zeta$, given by \cite{Fan,Qass}:
\begin{equation}\label{cond}
\zeta=\dfrac{i q^2}{4 \pi \hbar} \ln \left(     \dfrac{2 \mu_c-\mathcal{W} }{2 \mu_c+\mathcal{W}} \right)+\dfrac{iq^2K_B T}{\pi \hbar^2\mathcal{W}}   2\ln(e^{-\mu_c/K_B T}+1),     
\end{equation} 
where $\mathcal{W}=(\omega/2\pi+i \tau^{-1})$ with $\tau$ being the scattering relaxation time, $\mu_c=\hbar V_f \sqrt{\pi n_0}\sqrt{1+2 \mathcal{C} v_m/q \pi n_0 }$ represents the chemical potential of the graphene with $q$ is the electron charge, $n_0$ is the electron density, and $V_f$ denotes the Dirac fermions velocity. We consider the case of having an optical pump at $\omega_1$, besides two upper and lower side optical signals at $\omega_2$ and $\omega_3$, respectively. These optical fields are launched to the graphene layer as surface plasmon polariton modes. The interaction between these fields are enabled, by setting the microwave frequency equal $\omega_m=\omega_2-\omega_1=\omega_1-\omega_3$, through the electrical modulation of the graphene conductivity \cite{Qaass}.
To model the interaction between the microwave and the optical fields, for weak driving microwave signal, we expand the chemical potential of the graphene $\mu_c = \mu'_c + v \mu''_ce^{-i\omega_m t}+c.c $ to the first order in term of $v_m=$, where $\mu'_c=\hbar V_f \sqrt{\pi n_0}$ and $\mu''_c =\hbar V_f \mathcal{C}/q\sqrt{\pi n_0}$.
This expansion is obtained under the assumption of $\mathcal{C} v_m << q \pi n_0$.
 Then, the conductivity of the graphene in Eq.(\ref{cond}) is modified and can be written as $\zeta_c = \zeta'_c + v_m \zeta''_c e^{-i\omega_m t}+c.c $, where $\zeta'$ has the same value as given in Eq.(\ref{cond}) and $\zeta''=iq^2 [\mathcal{W}  \mu''_c /\pi\hbar [(2\mu'_c)^2-\mathcal{W}^2\hbar^2]+K_B T \mu''_c \tanh(\mu_c/2K_B T) /\mathcal{W}K_B T]$. Therefore, the effective permittivity of the graphene plasmonic waveguide is given by:
\begin{equation}
\epsilon_{eff}=\epsilon'+v \epsilon'' e^{-i\omega_m t}+c.c, \label{eqp}
\end{equation}
where $\epsilon'=(c \beta'/\omega)^2$, $\epsilon''=2 c^2 \beta' \beta''_j/\omega^2$ and $\beta'=\beta$. Here, $\beta''=\beta'\zeta'' \left(1- (Z_0\zeta'/2)^2\right)^{-1}/\zeta'$. Consequently, a simplified description of the interaction can be obtained by substituting the effective permittivity $\epsilon_{eff}$ ( Eq.(\ref{eqp})) in the governing classical Hamiltonian: 
\begin{equation}
H=\dfrac{1}{2}\mathcal{C} v^2 \mathcal{A}_r+\dfrac{1}{2}\int_{x,y,z}\left(\epsilon_0\epsilon_{eff}  |\vec{\mathcal{E}}_t|^2+\mu_0|\vec{\mathcal{H}}_t|^2  \right)dx dy dz,
\end{equation}
 where $\vec{\mathcal{E}}_t =\sum^3_{j=1} \vec{\mathcal{E}}_j $ is the total electric field with $\vec{\mathcal{E}}_j=\mathcal{U}_j(z)\Big( \mathcal{D}_{x_j}(x) \vec{e}_x
  +\mathcal{D}_{z_j}(x) \vec{e}_z \Big)e^{-i(\omega_j t-\beta_j z)}+c.c$ (for $j=1,2,3$) and $\vec{\mathcal{H}}_t$ is the total magnetic field. The corresponding quantized Hamiltonian that describes the three optical modes of frequencies $\omega_1$, $\omega_2$, and $\omega_3$, and the microwave mode, reads:
\begin{eqnarray}
\hat{H}=\hbar \omega_m  \hat{b}^\dagger \hat{b}&+&\hbar \sum_j \omega_j \hat{a}^ \dagger_i \hat{a}_j + \hbar  g_2 \hat{a}^\dagger_2 \hat{a}_1 \hat{b}\non
\\&+&\hbar g_3 \hat{a}^\dagger_3 \hat{a}_1 b^\dagger+h.c.,
\end{eqnarray}
where $\hat{a}_j=\mathcal{U}_j \sqrt{\xi_j\epsilon_0\epsilon'_{eff_j} V_L/\hbar \omega_j}$ is the annihilation operator of the \textit{j-th} optical mode, $\hat{b}=v\sqrt{\mathcal{C} \mathcal{A}_r /2 \hbar \omega_m}$ is the annihilation operator of the microwave mode, $V_L=\mathcal{A}_r \int (|\mathcal{D}_{xj}|^2+|\mathcal{D}_{zj}|^2  ) dx$, $\xi_j =\dfrac{1}{2}+\dfrac{\mathcal{A}_r \mu_0 \int |\mathcal{D}_{y_j}|^2 dx}{2 V_L\epsilon_0\epsilon'_{j}}$, and $\mathcal{A}_r=L\times W$ represents the cross sectional area of the capacitor.  The coupling factors $g_j$ are given by: 
\begin{eqnarray}
 g_j&=& \dfrac{\epsilon''_{j} l_{1j}}{2\sqrt{\xi_1\xi_j}}   \sqrt{\dfrac{2\omega_1\omega_j\hbar\omega_m}{\mathcal{C}\mathcal{A}_r \epsilon'_{1}\epsilon'_{j}    }}\text{sinc}\left(  \Theta_j \right) e^{i\Theta_j},
\end{eqnarray}
where $\Theta_j=(-1)^j\Big(\beta_1-(-1)^j\beta_j\Big)L/2$ and $l_{1j}=\dfrac{\int( \mathcal{D}^*_{x1}\mathcal{D}_{xj}       +\mathcal{D}^*_{z1}\mathcal{D}_{zj})dx}{\sqrt{\int (|\mathcal{D}_{x1}|^2+|\mathcal{D}_{z1}|^2) dx \int (|\mathcal{D}_{xj}|^2+|\mathcal{D}_{zj}|^2) dx}} $.
Here, $j=2,3,b$. We note here that the surface plasmon polariton mode at frequency $\omega_1$ is very intensive and can be treated classically. As a matter of fact, coupling to such intensive field provides the required gain to compensate the propagation losses of the quantum fields at frequencies $\omega_{2,3}$. 

By considering a rotating frame at frequency $\omega_j$, and introducing the corresponding noise terms according to the fluctuation-dissipation theorem, the Heisenberg-Langevin equations of the microwave and optical operators can be derived from the Hamiltonian in Eq. (4),  given by: 
\begin{eqnarray}
\dot{\hat{b}}&=&-\gamma_m \hat{b}-i\mathcal{G}_2 \hat{a}_2 -i\mathcal{G}_3 \hat{a}^\dagger_3 +\sqrt{2 \gamma_m}\, \hat{b}_{in},\non\\
\dot{\hat{a}}_2&=&-\gamma_2 \hat{a}_2 - i\mathcal{G}_2 \hat{b} +\sqrt{2 \gamma_2}\, \hat{a}_{in_2},\non\\
\dot{\hat{a}}_3&=&-\gamma_3 \hat{a}_3 - i\mathcal{G}_3 \hat{b}^\dagger +\sqrt{2 \gamma_3}, \, \hat{a}_{in_{3}},\label{eqm}
\end{eqnarray}
where $\gamma_m$ represents the damping rate of the microwave mode, and $\gamma_{j}=2 v_g Im(\beta_j)$ is the decay rate of the \textit{j-th} optical mode, and $v_g=\frac{\partial f}{\partial \beta}$ is the group velocity of the SPP mode. The operators $\hat{a}_{in_i}$ and $\hat{b}_{in}$ are zero-average input noise operators for the \textit{j-th} optical mode and the microwave mode, respectively. Those can be characterized by $\langle \hat{a}^\dagger_{in_j}(t), \hat{a}_{in_{j'}}(t')\rangle= n_ {j }\delta_{j ,{j'}} \delta (t-t')$ and $\langle \hat{b}^\dagger_{in}, \hat{b}_{in}\rangle= n_ {b } \delta (t-t')$ \cite{gard}. The mean thermal populations of the \textit{j-th} optical mode and the microwave mode are given by $n_{j} =( e^{\hbar \omega_{j} /k_B T }-1)^{-1}$ and $ n_{b} = (e^{\hbar \omega_{m} /k_B T }-1)^{-1} $, respectively, where $k_B $ is the Boltzmann constant. The optical thermal photon number can be assumed $n_j\approx0$ because of $\hbar \omega_j/K_b T\gg0$, whereas the microwave thermal photon number $n_m$ is significant and can not be conceived identical to \textit{zero} even at a very low temperature. 

\begin{figure}[h]
\centering
\includegraphics[width=0.99\columnwidth]{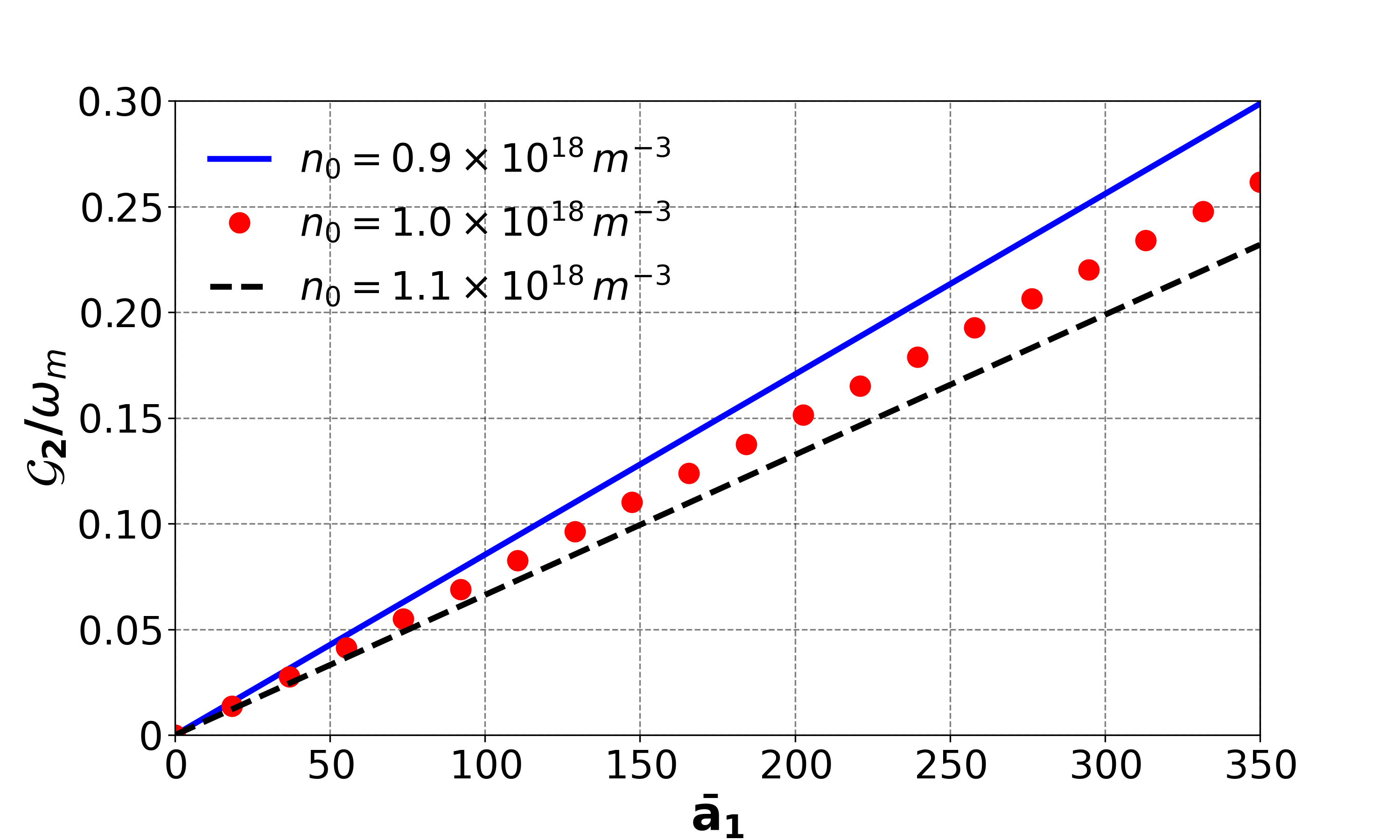}
\caption{The coupling strength versus the classical optical pump. Different doping concentration $n_0$ is considered. Here, $\frac{\omega_m}{2\pi}=10 GHz$, $\gamma_2=\gamma_3=2.6\times 10^{10} Hz$, and $\gamma_m=10^9 Hz$}. 
\label{fig112}
\end{figure}

The Eq.(\ref{eqm}) above shows that the microwave mode is coupled to the two optical modes $\hat{a}_2$ and $\hat{a}_3$ by the effective couplings $\mathcal{G}_{2}=\bar{a}_1 g_{2}$ and $\mathcal{G}_{3}=\bar{a}_1 g_{3}$, respectively, where $\bar{a}_1 $ being the classical amplitude operator of the optical pump at frequency $\omega_1$. The corresponding power of this optical pump is given by $p_1=\frac{1}{2} c \epsilon_0 n_1 \lvert\bar{a}_1\rvert^2 \big(\frac{\hbar \omega_1}{\xi\epsilon_0\epsilon'_{1} V_L}\big)
 S_m $, where $n_1=\sqrt{\epsilon'_{1}}$ and  $S_m=W \int_{\mathcal{-\infty}}^{+\infty} \big(\lvert \mathcal{D}_{x_1} \rvert\ ^2 +\lvert \mathcal{D}_{z_1} \rvert\ ^2 \big)\partial x$ are the refractive index and the transverse cross sectional area of the optical pump mode, respectively.
Importantly, the $\bar{a}_1 $ operator can be controlled to adjust the effective coupling and provide significant entanglement between the two optical modes $\hat{a}_2$ and $\hat{a}_3$. The feasible coupling values can be estimated by considering practical parameters. For instance, on considering a graphene layer of width $W=1 \mu m$, and a superconducting capacitor of $\mathcal{C}=13.3 \mu F/m^2$ $\big($ assuming a silicon filling material and a separation distance $d=1 \mu m$ $\big)$, the coupling strength is displayed in Fig.(\ref{fig112}) versus the classical operator $\bar{a}_1$. Here, $\mathcal{G}_{2} \approx \mathcal{G}_{3}$, the temperature is considered $20 mk$ for quantum microwave operation, $\tau=6 ps$, $V_f=10^6 m/s$, $\frac{\omega_1}{2\pi}=193 THz$, and $\omega_{2,3}=\omega_1\pm \omega_m$. We note that the considered $\bar{a}_1$ range in Fig.2, which is between $0$ and $350$, is corresponding to pump power $p_1$ range  between $0$ and $26 mW$. This is a practical power range that can be supported using a readily available laser source.

To gain more insight, we rewrite the two field operators $\hat{a}_2$ and $\hat{a}_3$ in term of the Bogolyubov operators $\hat{\mathcal{A}}_2=\hat{a}_3 \cosh r  +\hat{a}^\dagger_2 \sinh r $ and $\hat{\mathcal{A}}_3=\hat{a}_2 \cosh r +\hat{a}^\dagger_3 \sinh r $, where $\cosh r=\mathcal{G}_2/\mathcal{G}$, $\sinh r=\mathcal{G}_3/\mathcal{G}$ with $\mathcal{G}=\sqrt{\mathcal{G}^2_2-\mathcal{G}^2_3}$.
It then follows that the motion equations in Eq.(\ref {eqm})  can be presented in term of the Bogolyubov modes, given by:
\begin{eqnarray}
\dot{\hat{b}}&=&-\gamma_m \hat{b}-i\mathcal{G} \hat{\mathcal{A}}_3 +\sqrt{2 \gamma_m}\, \hat{b}_{in},\non\\
\dot{\hat{\mathcal{A}}}_2&=&-\gamma_2 \hat{\mathcal{A}}_2  +\sqrt{2 \gamma_2}\, \hat{\mathcal{A}}_{in,2},\non\\
\dot{\hat{\mathcal{A}}}_3&=&-\gamma_3 \hat{\mathcal{A}}_3 - i\mathcal{G} \hat{b} +\sqrt{2 \gamma_3} \, \hat{\mathcal{A}}_{in,3},\label{eqb}
\end{eqnarray}
where $\hat{\mathcal{A}}_{in,2}=\hat{a}_{in,3} \cosh r  +\hat{a}^\dagger_{in,2} \sinh r $ and $\hat{\mathcal{A}}_{in,3}=\hat{a}_{in,2} \cosh r +\hat{a}^\dagger_{in,3} \sinh r $. It can be infer from the above set of equations that the considered modes of the two optical fields are entangled\cite{Sbam, genes}. As will be elaborated in the next section. 

\section{Stationary Output Entanglement }\label{sec3}
Now we consider the problem of entanglement between the outgoing light fields of the optical modes $\hat{a}_2$ and $\hat{a}_3$. According to the input-output theory \cite{collett}, the output fields operators $\hat{a}_{out_2}$ and $\hat{a}_{out_3}$ are related to the two cavity operators $\hat{a}_2$ and $\hat{a}_3$ by $\hat{a}_{out_{2}}(t)=\sqrt{\gamma_{2}} \hat{a}_2(t) -\hat{a}_{in_{2}}(t) $ and $\hat{a}_{out_{3}}(t)=\sqrt{\gamma_3} \hat{a}_3(t) - \hat{a}_{in_{3}}(t) $, respectively.  To study the stationary entanglement between the output optical modes, it is convenient to rewrite Eq.(\ref{eqm}) in  the following compact matrix form:
\begin{equation}
\dfrac{d}{dt} \mathcal{R}(t)=  \mathcal{A} \mathcal{R} (t)+\mathcal{D} \mathcal{R}_{in}(t), \label{eqc}
\end{equation}
where $\mathcal{R}^T =\{\hat{a}_{2}, \hat{a}^\dagger_{2}, \hat{a}_{3}, \hat{a}^\dagger_{3}, \hat{b}, \hat{b}^\dagger \}$ is the column vector of the field operators, $\mathcal{R}^T_{in} =\{\hat{a}_{in, 2}, \hat{a}^\dagger_{in,2}, \hat{a}_{in, 3}, \hat{a}^\dagger_{in,3}, \hat{b}_{in}, \hat{b}^\dagger_{in} \}$ is the column vector of the corresponding noise operators, and the superscript $T$ indicating transposition. Here, $\mathcal{A}$ is the drift matrix with elements that can be easily obtained from the Langevin equations set in Eq.(\ref{eqm}), $\mathcal{D}$ is the coefficients matrix of the corresponding input noise operators. For a drift matrix $\mathcal{A}$ with eigenvalues in the left half of the complex plane, the interaction is stable and reaching the steady-state. The solution can be obtained in the frequency domain, by applying Fourier transform to Eq.(\ref{eqc}), given by:
\begin{equation}
\mathcal{R}_{out}(\omega)=\left[  \mathcal{F} \mathcal{M}(\omega) \mathcal{D}  - {\bf \mathrm{I}} \right] \mathcal{R}_{in},\label{eqo}
\end{equation}
where $\mathcal{R}^T_{out}(\omega)=\{ \hat{a}_{ out_2} (\omega), \hat{a}^\dagger_{out_2}(\omega), \hat{a}_{out_3}(\omega), \\ \hat{a}^\dagger_{out_3} (\omega), \hat{b}(\omega), \hat{b}^\dagger(\omega) \}$, $\mathcal{R}_{in}(\omega)$ is the Fourier transform of $\mathcal{R}_{in}(t)$, $ \mathcal{M}(\omega)=[\mathcal{A}-i\omega]^{-1} $, 
 $\mathcal{F}(\omega)= \text{diag} \{  \sqrt{2 \gamma_2},\sqrt{2 \gamma_2} ,\sqrt{2 \gamma_3} ,\sqrt{2 \gamma_3}, 1 , 1  \}$ with  {\bf $\mathrm{I}$} is $6 \times 6$  identity matrix. Given that the operators of the quantum input noise are Gaussian, the steady-state of the system is completely described by first and second-order moments of the output field operators. In particular, it is convenient to introduce the quadratures $\hat{X}_{out_2}=(\hat{a}_{out_2}+\hat{a}^\dagger_{out_2})/\sqrt{2} $, $\hat{Y}_{out_2}=(\hat{a}_{out_2}-\hat{a}^\dagger_{out_2})/\sqrt{2} i$,  $\hat{X}_{out_3}=(\hat{a}_{out_3}+a^\dagger_{out_3})/\sqrt{2} $, $\hat{Y}_{out_3}=(\hat{a}_{out_3}-\hat{a}^\dagger_{out_3})/\sqrt{2} i$, $\hat{X}_b=(\hat{b}+\hat{b}^\dagger)/\sqrt{2} $ and  $\hat{Y}_{b}=(\hat{b}-\hat{b}^\dagger)/\sqrt{2} i$. Then the correlation matrix (CM) $\mathcal{V}$ of the system is defined as $\mathcal{V}_{i j}=\langle  u_i u_j+u_j u_i  \rangle/2 $ where $u^T=\{ \hat{X}_{out_2}, \hat{Y}_{out_2}, \hat{X}_{out_3}, \hat{X}_{out_3}, \hat{X}_{b}, \hat{Y}_{b}\}$ is the vector of the quadratures for the output modes. From Eq.(\ref{eqo}), the stationary solution for the covariance matrix $\mathcal{V}$ of the output modes can be obtained by:
\begin{eqnarray}
\mathcal{V}=\int^\infty_{-\infty}   \mathcal{Q} \mathcal T(\omega) \mathcal{N}  T(-\omega)^T \mathcal{Q}^T  d\omega  ,
\end{eqnarray}
where $\mathcal{Q} =\text{diag} \{ \mathcal{Q}_2, \mathcal{Q}_3, \mathcal{Q}_b\}$, $\mathcal{Q}_j=\dfrac{1}{\sqrt{2}}\begin{pmatrix}  1 & 1 \\
-i & i  \end{pmatrix}$,  $T(\omega) = \mathcal{F} (\omega)\mathcal{M}(\omega) \mathcal{D}  - {\bf I} $ and $\mathcal{N}=\text{diag} \{ \mathcal{N}_2, \mathcal{N}_3, \mathcal{N}_b\}$ is the diffusion matrix. Here, $\mathcal{N}_b=\begin{pmatrix}  0 & n_m+1\\
 n_m& 0  \end{pmatrix}$, and $\mathcal{N}_j$ stands for $2\times2$ matrix of $\{\mathcal{N}_j\}_{12}=1$ (for j=2,3) while all other elements are being zero. 

We now consider the generation of stationary output entanglement between the two optical modes $\hat{a}_{out_2}$ and $\hat{a}_{out_3}$. The covariance matrix for $\hat{a}_{out_2}$ and $\hat{a}_{out_3}$ can be introduced as in the following block form:
\begin{equation}\label{eqv}
v=\begin{pmatrix}
\mathcal{V}_{a_{2}} &  \mathcal{V}_{a_{23}} \\
\mathcal{V}^T_{a_{23}} &  \mathcal{V}_{a_3} \\
\end{pmatrix},
\end{equation}
with $\mathcal{V}_{a_2}$ and $\mathcal{V}_{a_3} $ are $2 \times 2$ covariance matrices for the two ouput optical $\hat{a}_{out_2}$ and $\hat{a}_{out_3}$ modes, respectively. The correlation between $\hat{a}_{out2}$ and $\hat{a}_{out3}$ can be described by the $2 \times 2$\, $ V_{a_{23}}$ matrix. The stationary entanglement between Alice's (mode $\hat{a}_{out_2}$) and Bob's (mode $\hat{a}_{out_3}$) can be measured by the negativity \cite{vidal} (i.e., quantified by the logarithmic negativity \cite{adessoo,eisert,plenio}):
\begin{equation}\label{En}
E_{\mathcal{N}}=\max[0, -\ln 2 \eta^-],
\end{equation}
where $\resizebox{.6\hsize}{!}{$ \eta^-=2^{-1/2}  \sqrt{\sum^2  (\mathcal{V}) -\sqrt{\sum^2 ( \mathcal{V}) - 4 \det( \mathcal{V})}} $} $ is the smallest symplectic eigenvalue of the partially transposed $4 \times 4$ covariance matrix (CM) $v$ with$\sum (\mathcal{V})=\det(\mathcal{V}_{a_2})+\det(\mathcal{V}_{a_{3}})-2 \det(\mathcal{V}_{a_{23}})$. 
\begin{figure}[tb]
\includegraphics[width=1.02\columnwidth]{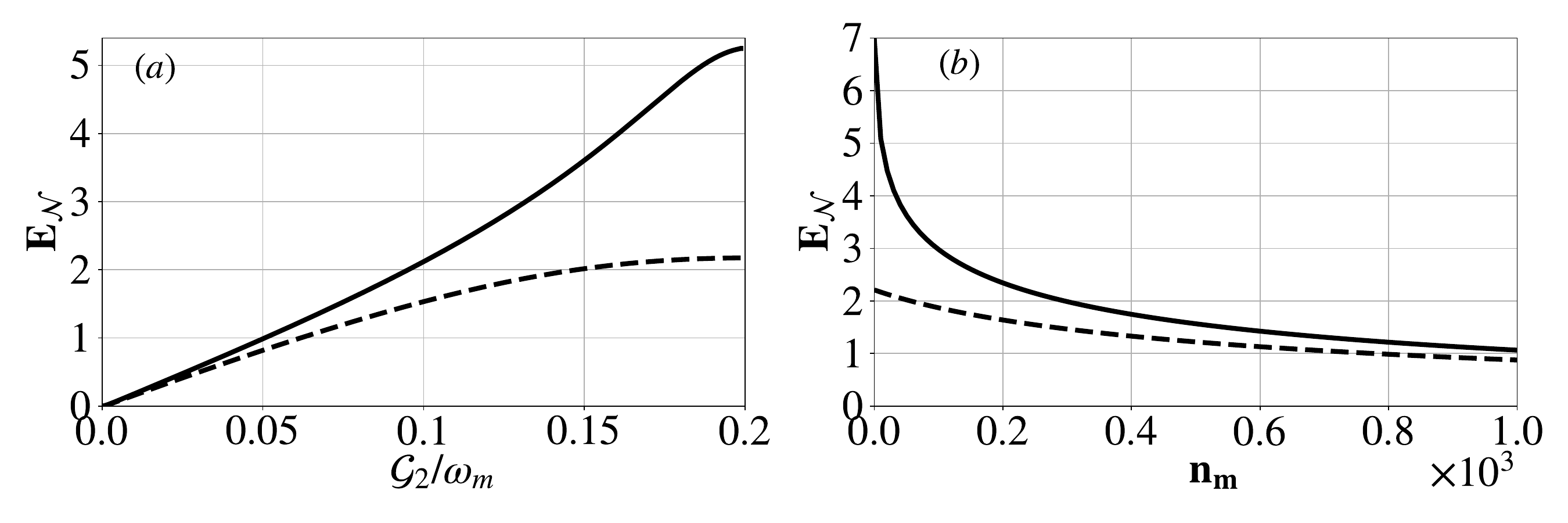}
\caption{(a) Logarithmic negativity $E_{\mathcal{N}}$ measure of entanglement between Alice and Bob modes plotted as function of $\mathcal{G}_2$ coupling for  $G_3=0.2\omega_m$ and $n_m=10$. (b) $E_{\mathcal{N}}$ is plotted as function of the microwave thermal excitation $n_m$ for $\mathcal{G}_2=0.18$, $\mathcal{G}_3=0.2\,\omega_m$. Here, the sold black curves are in the absence of optical losses while black dashed curves refer to the situation when optical losses are considered, with attenuation $\Gamma=0.005\, \mathrm{dB/km}$, distance $L=20 \mathrm{km}$, and $\eta_0=0.9$. The other parameters are $\mathcal{G}_2 \approx  \mathcal{G}_3=0.2\,\omega_m$,  $\gamma_2=\gamma_3= 0.001\,\omega_m $ and $\gamma_m=0.02\,\omega_m$.}
\label{fig1}
\end{figure}
A non-zero value of $E_\mathcal{N}$ quantifies the degree of entanglement between Alice's and Bob's modes. 
When the two optical output modes possess Einstein-Podolsky-Rosen (EPR) correlations, they can be immediately exploited for transfer of the quantum information over long distances (\textit{i.e.}, teleportation of an unknown coherent state). On the other hand, in such a case of having a long-distance quantum communication, it is important to consider the optical losses for the two optical fields that are traveling in a classical channel (e.g., a free space or an optical fiber). These optical losses can be modeled as a beam splitter with transmissivity $\eta=\eta_0 e^{-\Gamma L/10}$ \cite{Barbosa}, where $L$ is the traveled distance by two optical fields (assumed to be equal for both), $\Gamma$ describes the attenuation of the two fields in free space in $\mathrm{dB/km}$, and $\eta_0$ accounts for all possible inefficiencies \cite{asjad, asjad2}. In the presence of these losses and inefficiencies, the covariance matrix in Eq.(\ref{eqv}) becomes $v_{loss}=\eta \, v+(1-\eta) {\bf I}/2 $, where  {\bf I} is $4 \times 4$ identity matrix. Here, we consider resonance condition for the two optical modes.  In Fig.(\ref{fig1}), we plot the Logarithmic negativity $E_{\mathcal{N}}$ to quantify the entanglement between the output modes against the effective coupling $\mathcal{G}_2$ and the microwave thermal photon $n_m$ in the absence (black solid curves) and presence (dashed black curve) of path losses. The black solid curves are with no optical loss, i.e, $\eta=1$, while the black dashed curves refer to the situation when optical losses are considered with detection efficiency $\eta_0=0.9$,  $L=20 \mathrm{km}$, and an attenuation of $\Gamma=0.005 \mathrm{{dB/km}}$ for a free space channel of a clear day without any turbulence \cite{Fischer}. In Fig.\ref{fig1}(a), we plot $E_{\mathcal{N}}$ as function of the parameter $\mathcal{G}_2/\omega_m$ while the coupling $\mathcal{G}_3$ is fixed at $\mathcal{G}_3=0.2\omega_m$. We observe that the maximum value of entanglement between the two output optical modes is achieved when the two couplings fulfill the condition $\mathcal{G}_2\approx \mathcal{G}_3$. This can be explained by noting that the squeezing parameter $r$ in Eq.(\ref{eqb}), which is defined as the ratio of $\mathcal{G}_3$ and $\mathcal{G}_2$, is approaching one ($r=\mathcal{G}_3/\mathcal{G}_2\approx 1$) at this condition. In Fig. \ref{fig1}(b), we also study the robustness of the steady state entanglement between the two optical modes $\hat{a}_{out,2}$ and $\hat{a}_{out,3}$ as function of the microwave thermal population $n_{m}$ at the optimal condition of $\mathcal{G}_2 \approx\mathcal{G}_3$. One can observe from Fig. \ref{fig1}(b) that the proposed entanglement is robust against the microwave thermal population. Here, we note that the condition $\mathcal{G}_2\approx\mathcal{G}_3$ can be obtained by controlling the graphene properties including the doping concentration and the layer dimensions. 

\section{ Continuous variable Teleportation}\label{sec4}
The EPR-like continuous variable entanglement generated between the two output fields can be characterized in term of its effectiveness as a quantum channel for quantum teleportation. The performance of the quantum channel can be realized in term of the teleportation fidelity of an unknown coherent state between two distant nodes labeled as Alice's and Bob's, as shown in Fig.\ref{fig11}.\\ 
The teleportation process can be summarized as in the following four steps:
\begin{figure}[h]
\centering
\includegraphics[width=0.99\columnwidth]{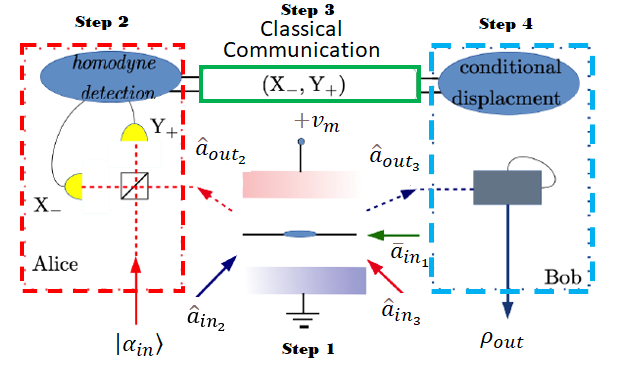}
\caption{ The proposed scheme for quantum teleportation of the light signals. Alice (sender) and  Bob (receiver) share EPR quantum channel given by an EPR pair optical modes. Alice uses a beam splitter to mix the part of entangled state in her hand with an unknown coherent state which is to be teleported.   She performs the Bell measurement. Alice then sends the results of her Bell's measurements to Bob. Bob performs a specific displacement operation on Alice's sent measurements to obtain the teleported state.}
\label{fig11}
\end{figure} 
\begin{enumerate}
\item The two output optical fields $\hat{a}_{out_2}$ and $\hat{a}_{out_3}$ are entangled via the proposed microwave enabled plasmonic graphene waveguide and consequently possess EPR correlation. The two fields are propagating to Alice and Bob, respectively. Hence, the two optical fields form a quantum channel of two-mode Gaussian entangled state $\rho_{23}$ (this step corresponds to the fields co-propagating along the graphene waveguide in the middle of Fig.\ref{fig11}).
\item Second, Alice combines an unknown input coherent state $|\alpha_{in}\rangle$, that is to be teleported,  with the part of the entangled state in her hand (i.e., $a_{out_2}$ mode),  using a beam splitter and a two sets of homodyne detectors, to measure the amplitudes $X_-=\sqrt{2} (\hat{X}_{in}-\hat{X}_{out_2})$ and $Y_+=1/i \sqrt{2}(\hat{Y}_{in}+\hat{Y}_{out_3})$. Here $X_{in}$ and $Y_{in}$ are the quadrature of input state (this step corresponds to the red dashed square in Fig.\ref{fig11}).
\item Third, the measurement outcomes of Alice are sent to Bob via classical channel (this step corresponds to the green square in Fig.\ref{fig11}).
\item Finally, Bob performs a conditional displacement on his mode $a_{out_3}$ according to the measurement outcomes and constructs the output state $\rho_{out}$ (this final step corresponds to the blue dashed square in Fig.\ref{fig11}).
\end{enumerate}
To asses the quality of the teleportation, the overlap between the input and the output states can be calculated  using the concept of teleportation fidelity, defined by $F = |\langle \alpha_{in}|  \rho_{out}  |\alpha_{in}\rangle$. Since the system is Gaussian, the teleportation fidelity can be characterized in terms of the Gaussian characteristic functions of the quantum channel and the input coherent state, given respectively by:
\begin{equation}
\chi_{ch}(\vec{\alpha}_{2},\vec{\alpha}_{3} )= \exp\left[ 
 \begin{pmatrix}
  \vec{\alpha}^T_{2} & \vec{\alpha}^T_{3}
  \end{pmatrix}  v  \begin{pmatrix}
  \vec{\alpha}_{2}\\
  \vec{\alpha}_{3},
  \end{pmatrix} + i \vec{d}_{ch} 
  \begin{pmatrix}
  \vec{\alpha}_{2}\\
  \vec{\alpha}_{3} \end{pmatrix}  \right],
\end{equation}
and
\begin{equation}
\chi_{in}(\vec{\alpha}_{in} )= \exp\left[ 
  \vec{\alpha}^T_{in} \mathcal{V}_{in} \vec{\alpha}_{in},
  + i \vec{d}_{in} \alpha_{in}  \right],
  \end{equation}
where $\vec{\alpha}_{\imath}=(\Im [\alpha_{\imath}],-\Re[\alpha_{\imath}])$ for $\imath=2,3,in$ is a two-dimensional vector corresponding to the complex phase-space variable $\alpha_{\imath}$, and $\vec{d}_{\jmath} $ $\left(\jmath=ch, in\right)$  is a drift vector. Here, $\mathcal{V}$ is the covariance matrix for the channel given in Eq.(\ref{eqv}), and $\mathcal{V}_{in}=\dfrac{1}{2}\text{diag}(1, 1)$ is the covariance matrix of the input coherent state. Consequently, the expression of the teleportation fidelity $F$  can be written as \cite{pirandola2004}:
\begin{equation}\label{tf}
F=\dfrac{1}{\pi}\int e^{\mu \alpha^*-\mu^* \alpha}  |\chi_{in}(\alpha)|^2  [\chi_{ch}(-\alpha^*,\alpha ) ]^*d^2 \alpha, 
\end{equation}
where $\mu$ is the displacement that Bob needs to perform on his side to cancel the effects of the channel displacement  $d_{ch}$. Therefore, when Bob chooses the value of the additional displacement $\mu$ to be exactly balancing $d_{ch}$, the teleportation Fidelity $F$ in Eq.(\ref{tf}) can be simplified to:  
 \begin{eqnarray}
 F &=&\dfrac{1}{\sqrt{\det {\Gamma}}},\,\,\,\, \non\\
 \Gamma&=&2\mathcal{V}_{in} +  \mathcal{Z} \mathcal{V}_{a_2} \mathcal{Z} + \mathcal{V}_{a_3}  -  \mathcal{Z}  \mathcal{V}_{a_{23}} - \mathcal{V}_{a_{23}}^T \mathcal{Z} , \non
 \end{eqnarray}
 where $\mathcal{Z} =\text{diag}(1,-1)$.
\begin{figure}[tb]
\includegraphics[width=1\columnwidth]{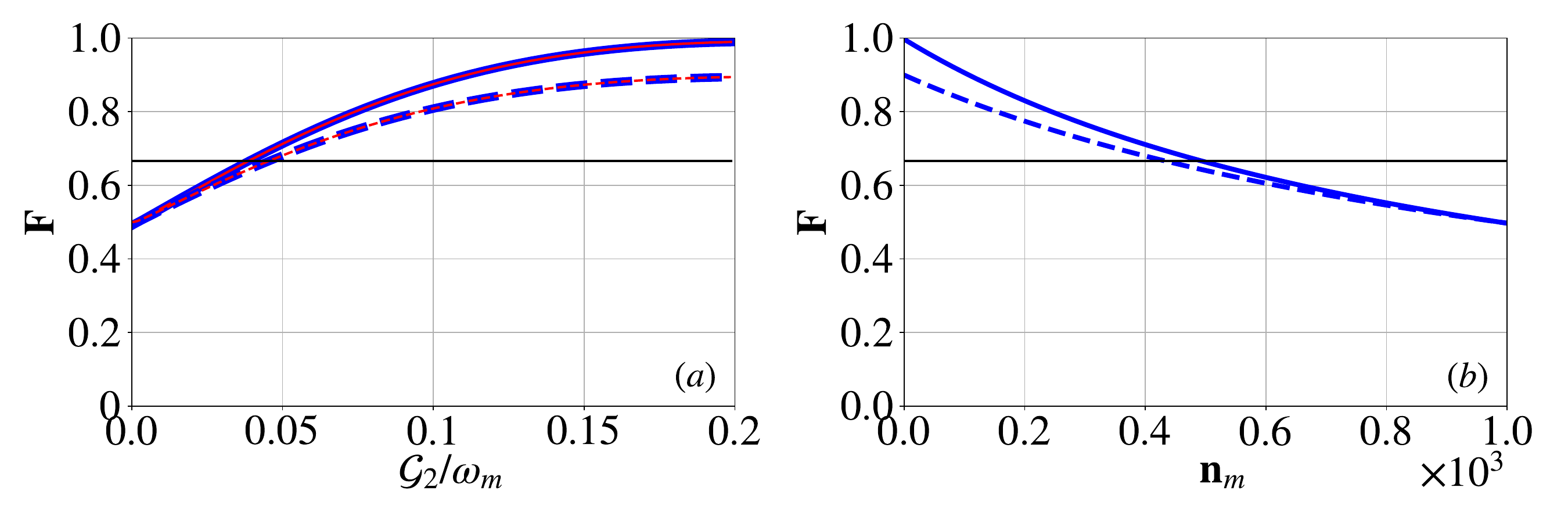}
\caption{(a) Teleportation fidelity (black curve) $F$ of the coherent state as a function of the coupling $\mathcal{G}_2/\omega_m$ at fixed value of $n_m=10$ and $\mathcal{G}_3=0.2\omega_m$.  (b) Teleportation fidelity $F$ as a function of $n_m$ at $\mathcal{G}_2=0.18\omega_m $. The solid  red curve is for the upper bound of the teleportation fidelity. Here, the solid blue curves are without optical losses while dashed blue curves correspond to the case with optical losses. The solid (dashed) red curve is for the upper bound of the teleportation fidelity in the absence (presence) of the optical losses.} The black horizontal line represents the threshold for secure teleportation $F_{thr}=2/3$. The other parameters are same as in Fig.\ref{fig1}.
\label{fig2}
\end{figure} 
Moreover, the upper bound set by the entanglement on the fidelity of the CV teleportation, optimized over the local operations, is given by \cite{mari}:
\begin{equation}\label{eb}
 F^{opt} =\dfrac{1}{1+\exp^{-E_{\mathcal{N}}}},
\end{equation}
where $E_{\mathcal{N}}$ is the logarithmic negativity of the two-mode entanglement shared between Alice and Bob. In our case, the entangled resource shared by Alice and Bob is a CV Gaussian state with zero mean $\vec{d}_{ch}=0$. It then follows that the additional displacement that Bob is performing on his side to balance the channel displacement is also zero (i.e., $\mu=0$).

The corresponding quantum teleportation of an unknown optical coherent state using the obtained squeezed-state entanglement is shown in Fig.\ref{fig2}. In Fig.\ref{fig2}(a), we plot the teleportation fidelity of a coherent state as function of the effective coupling $\mathcal{G}_2$, considering the microwave thermal population $n_m=10$ and having the coupling $\mathcal{G}_3= 0.2\omega_m$ in the absence (solid blue curves) and presence (dashed blue curves) of optical path losses.  It can be seen from Fig.\ref{fig2}(a) that the maximum value of the fidelity (blue curve) is achieved when $\mathcal{G}_2=0.18 \omega_m$ and $\mathcal{G}_3=0.2\omega_m$. This is the same condition obtained for the optimal entanglement in Fig.\ref{fig1}(a).
Moreover, we observed that the maximum value of the fidelity (red curves) adheres the upper bound, defined in Eq.(\ref{eb}). In Fig.\ref{fig2}(b), we also study the teleportation fidelity for an unknown coherent state as function of the microwave thermal excitation $n_m$ while considering $\mathcal{G}_3=0.2\omega_m$ and $\mathcal{G}_2=0.18\omega_m$. Interestingly, it is found that the proposed teleportation is very robust against the microwave thermal population and optical losses. For example, the teleportation fidelity is above $2/3$ even for $n_m\approx500$ and $\eta_0=0.9$ at a distance $L=20 \mathrm{km}$ with $\Gamma =0.005 \mathrm{dB/km}$ for clean day in free space. In low noise fiber optics this is equalvent to $7 \mathrm{km}$ with attenuation $0.16 \mathrm{dB/km}$ \cite{Afzelius}.This is a realization of quantum teleportation of an unknown coherent state $|\alpha\rangle$ entering the device as Alice and is being teleported to Bob. It is worth mentioning that for secure quantum teleportation of coherent state, a fidelity grater then the threshold fidelity $F_{thr}=2/3$ is required, which is unreachable without the use of entanglement.
\\\\

\section{Conclusion} \label{sec5}
We have presented a scheme for generating a continuous variable two-mode squeezed entangled state between two optical fields independent of each other in a hybrid optical-microwave plasmonic graphene waveguide system. We have further explored the two-mode squeezed entangled state between the two optical fields to demonstrate quantum teleportation of an unknown coherent state between two spatially distant nodes.  The achieved quantum teleportation is secure due to the fact that the fidelity  $F$  is above the threshold $F_{thr}=2/3$. We show that the continuous-variable entanglement (teleportation fidelity) can be controlled and enhanced through the interaction of the microwave mode with the two optical modes. Such pairs of entangled modes, combined with the technique of entanglement swapping, can be used as a quantum channel to teleport quantum state over large distances. 
\\\\
\noindent\textbf{Acknowledgments -} This research is supported by Abu Dhabi Award for Research Excellence under ASPIRE/Advanced Technology Research Council (AARE19-062) 2019.

\end{document}